\documentclass[aps,showpacs,twocolumn]{revtex4}

\usepackage{latexsym}
\usepackage{epsfig}
\usepackage{amssymb}

\newcommand{\R}{\mathcal{R}}

\begin{document}

\title{Metric-Palatini gravity unifying local constraints and late-time cosmic acceleration}

\author{Tiberiu Harko$^1$}\email{harko@hkucc.hku.hk}
\author{Tomi S. Koivisto$^{2,3}$}\email{tomi.koivisto@fys.uio.no}
\author{Francisco S.~N.~Lobo$^{4}$}\email{flobo@cii.fc.ul.pt}
\author{Gonzalo J. Olmo$^{5}$}\email{gonzalo.olmo@csic.es}

\affiliation{$^1$Department of Physics and Center for Theoretical
and Computational Physics, The University of Hong Kong, Pok Fu Lam Road, Hong Kong}
\affiliation{$^2$Institute for Theoretical Physics and Spinoza Institute, Utrecht University,
Leuvenlaan 4, 3584 CE Utrecht, The Netherlands}
\affiliation{$^{3}$  Institute of Theoretical Astrophysics, University of
  Oslo, P.O.\ Box 1029 Blindern, N-0315 Oslo, Norway}
\affiliation{$^4$Centro de Astronomia e Astrof\'{\i}sica da Universidade de Lisboa, Campo Grande, Ed. C8 1749-016 Lisboa, Portugal}
\affiliation{$^5$Departamento de F\'{i}sica Te\'{o}rica and IFIC, Centro Mixto Universidad de
Valencia - CSIC. Universidad de Valencia, Burjassot-46100, Valencia, Spain}

\date{\today}

\begin{abstract}
We present a novel approach to modified theories of gravity that consists of adding to the 
Einstein-Hilbert Lagrangian an $f(\R)$ term constructed \`{a} la Palatini. Using the 
respective dynamically equivalent scalar-tensor representation, we show that the theory can 
pass the Solar System observational constraints even if the scalar field is very light. This 
implies the existence of a long-range scalar field, which is able to modify the cosmological 
and galactic dynamics, but leaves the Solar System unaffected. We also verify the absence of 
instabilities in perturbations and provide explicit models which are consistent with local 
tests and lead to the late-time cosmic acceleration.
\end{abstract}

\pacs{04.50.Kd,04.20.Cv}

\maketitle

\section{ Introduction}

In the last ten years modified theories of gravity in which the Einstein-Hilbert Lagrangian is
supplemented with additional curvature terms, the so-called $f(R)$ theories \cite{fRgravity}, 
have been extensively studied in cosmology. More recently, generalizations of $f(R)$ gravity 
have been explored, namely, C-theories \cite{Amendola:2010bk}, and nonminimal curvature-matter 
couplings \cite{Harko:2010mv}. There was the hope that the addition of new curvature terms
could have an effect on the late-time cosmic dynamics \cite{expansion}, thus providing a
gravitational mechanism to explain the accelerated cosmic expansion rate. Though many of the 
initially proposed models naturally produced the desired late-time acceleration 
\cite{Carroll:2003wy}, it was soon observed that they were generically affected by serious 
problems. In particular, in the usual metric variational approach, the modified dynamics of 
$f(R)$ theories can be interpreted as due to a Brans-Dicke scalar field with parameter $w=0$ 
and a non-trivial potential $V(\phi)$. To satisfy the constraints imposed by laboratory and 
Solar System tests, a perturbative approach indicates that the $w=0$ scalar should be massive, 
with an interaction range not exceeding a few millimeters. Such scalars, obviously, cannot 
have any impact on the cosmology. Metric $f(R)$ theories, therefore, could only have a chance 
of being viable if by means of non-perturbative effects the scalar somehow managed to hide 
itself in local experiments while behaving as a long-range field at cosmic scales. Such models 
are also strongly constrained by the observation of cosmological perturbations and none of 
them seems to perform better than general relativity (GR) with a cosmological constant.

On the other hand, $f(R)$ theories have also been studied in the Palatini approach, where the
metric and the connection are regarded as independent fields \cite{Olmo:2011uz}. In this case, 
the gravitational dynamics is equivalent to a $w=-3/2$ Brans-Dicke theory with the same 
potential $V(\phi)$ as in the metric formulation. The $w=-3/2$ theory is characterized by a 
non-dynamical scalar, which makes it completely different from the other $w\neq -3/2$ theories 
in which the scalar is dynamical and, therefore, propagates. The non-dynamical nature of the 
$w=-3/2$ scalar implies that in vacuum the theory turns into general relativity (GR) with an 
effective cosmological constant $\Lambda_{\rm eff}$. This property guarantees the existence of 
accelerating de Sitter solutions at late times if $\Lambda_{\rm eff}$ is small. Despite this 
appealing property, all the Palatini $f(R)$ models studied so far with a small $\Lambda_{\rm 
eff}$ lead to microscopic matter instabilities and to unacceptable features in the evolution 
patterns of cosmological perturbations \cite{Olmo:2011uz,palatini_lss}.

In this work we present a new class of modified theories of gravity in which the usual
Einstein-Hilbert Lagrangian is supplemented with an $f(\R)$ Palatini correction. This type
of {\it hybrid} theory generically arises when perturbative quantization methods are considered
on Palatini backgrounds \cite{Flanagan:2003iw} which, on the other hand, have interesting
connections with non-perturbative quantum geometries \cite{Olmo:2008nf}. 

Already in classical gravitation, one has to specify two connections with physically distinct roles \cite{Amendola:2010bk}, 
so it is natural to consider that the action depends upon both of the associated curvatures. 
Metric-Palatini theories admit a non-standard scalar-tensor representation in terms of a 
dynamical scalar that needs not be massive to pass laboratory and Solar System tests. 
Microscopic matter instabilities are also absent in this model because the field is very 
weakly coupled to matter. In this theory, therefore, the scalar can play an active role in 
cosmology without being in conflict with local experiments. We provide explicit examples that 
illustrate these aspects. \\

\section{ Scalar-tensor representation of metric-Palatini gravity} 

Consider the action
\begin{equation} \label{eq:S_hybrid}
S= \frac{1}{2\kappa^2}\int d^4 x \sqrt{-g} \left[ R + f(\R)\right] +S_m \ ,
\end{equation}
where $S_m$ is the matter action, $\kappa^2\equiv 8\pi G$, $R$ is the Einstein-Hilbert term, 
$\R \equiv  g^{\mu\nu}\R_{\mu\nu} $ is the Palatini curvature, and $\R_{\mu\nu}$ is defined in 
terms of an independent connection $\hat{\Gamma}^\alpha_{\mu\nu}$  as
$\R_{\mu\nu} \equiv \hat{\Gamma}^\alpha_{\mu\nu , \alpha} - \hat{\Gamma}^\alpha_{\mu\alpha , 
\nu} + \hat{\Gamma}^\alpha_{\alpha\lambda}\hat{\Gamma}^\lambda_{\mu\nu} 
-\hat{\Gamma}^\alpha_{\mu\lambda}\hat{\Gamma}^\lambda_{\alpha\nu}$.

The action (\ref{eq:S_hybrid}) can be turned into that of a scalar-tensor theory by 
introducing an auxiliary field $A$ such that
\begin{equation} \label{eq:S_scalar0}
S= \frac{1}{2\kappa^2}\int d^4 x \sqrt{-g} \left[ R + f(A)+f_A(\R-A)\right] +S_m \ ,
\end{equation}
where $f_A\equiv df/dA$. Rearranging the terms and defining $\phi\equiv f_A$, $V(\phi)=A f_A-
f(A)$, the action (\ref{eq:S_scalar0}) becomes
\begin{equation} \label{eq:S_scalar1}
S= \frac{1}{2\kappa^2}\int d^4 x \sqrt{-g} \left[ R + \phi\R-V(\phi)\right] +S_m \ .
\end{equation}

Variation of this action with respect to the metric, the scalar $\phi$, and the connection 
leads to
\begin{eqnarray}
R_{\mu\nu}+\phi \R_{\mu\nu}-\frac{1}{2}\left(R+\phi\R-V\right)g_{\mu\nu}&=&\kappa^2 T_{\mu\nu} 
\ ,
\label{eq:var-gab}\\
\R-V_\phi&=&0 \,,\label{eq:var-phi}\\
\hat{\nabla}_\alpha\left(\sqrt{-g}\phi g^{\mu\nu}\right)&=&0 \,,\label{eq:connection}\
\end{eqnarray}
respectively.
The solution of Eq.~(\ref{eq:connection}) implies that the independent connection is the Levi-
Civita connection of a metric $t_{\mu\nu}=\phi g_{\mu\nu}$. This means that $\R_{\mu\nu}$ and 
$R_{\mu\nu}$ are related by
\begin{equation}
\R_{\mu\nu}=R_{\mu\nu}+\frac{3}{2\phi^2}\partial_\mu \phi \partial_\nu \phi-\frac{1}
{\phi}\left(\nabla_\mu \nabla_\nu \phi+\frac{1}{2}g_{\mu\nu}\Box\phi\right) \ ,
\end{equation}
 which can be 
used in Eq.~(\ref{eq:S_scalar1}) to obtain the following scalar-tensor theory
\begin{equation} \label{eq:S_scalar2}
S=\int \frac{d^4 x \sqrt{-g} }{2\kappa^2}\left[ (1+\phi)R +\frac{3}{2\phi}\partial_\mu \phi
\partial^\mu \phi -V(\phi)\right]+S_m .
\end{equation}
This action differs from the $w=-3/2$ Brans-Dicke theory in the coupling of the scalar to the
curvature, which in the $w=-3/2$ theory is of the form $\phi R$. As we will see, this simple 
modification will have important physical consequences. With the expression for $\R_{\mu\nu}$ 
and Eq.~(\ref{eq:var-phi}), Eq.~(\ref{eq:var-gab}) can be written as
\begin{eqnarray}
(1+\phi) R_{\mu\nu}&=&\kappa^2\left(T_{\mu\nu}-\frac{1}{2}g_{\mu\nu} T\right)+\frac{1}
{2}g_{\mu\nu}\left(V+\Box\phi\right)\nonumber \\&+&\nabla_\mu\nabla_\nu\phi-\frac{3}
{2\phi}\partial_\mu \phi \partial_\nu \phi \ \label{eq:evol-gab} .
\end{eqnarray}
The scalar field equation can be manipulated in two different ways that illustrate how this 
theory is related with the $w=0$ and $w=-3/2$ Brans-Dicke theories. Contracting Eq. 
(\ref{eq:var-gab}) with $g^{\mu\nu}$ and using Eq. (\ref{eq:var-phi}) we find
\begin{equation}\label{eq:phi(X)}
2V-\phi V_\phi=\kappa^2T+R \ .
\end{equation}
Similarly as in the Palatini ($w=-3/2$) case, Eq.~(\ref{eq:phi(X)}) tells us that $\phi$ can 
be expressed as an algebraic function of the scalar $X\equiv \kappa^2T+R$, i.e., 
$\phi=\phi(X)$. In the pure Palatini case, however, $\phi$ is just a function of $T$. The 
right-hand side of Eq.~(\ref{eq:evol-gab}), therefore, besides containing new matter terms 
associated with the trace $T$ and its derivatives, also contains the curvature $R$  and its 
derivatives. Thus, this theory can be seen as a higher-derivative theory in both the matter 
fields and the metric. However, an alternative interpretation without higher-order derivatives 
is possible if $R$ is replaced in Eq.~(\ref{eq:phi(X)}) with the relation
$R=\R+\left(3/\phi\right)\Box \phi-\left(3/2\phi^2\right)\partial_\mu \phi \partial^\mu \phi$, 
together with $\R=V_\phi$. One then finds that the scalar field is governed by the second-
order evolution equation
\begin{equation}\label{eq:evol-phi}
-\Box\phi+\frac{1}{2\phi}\partial_\mu \phi \partial^\mu \phi +\frac{\phi[2V-(1+\phi)V_\phi]}
{3}=\frac{\phi\kappa^2}{3}T \ .
\end{equation}
This latter expression shows that, unlike in the Palatini case \cite{Olmo:2011uz}, the scalar 
field is dynamical and not affected by the microscopic instabilities found in Palatini models 
with infrared corrections.\\

\section{ Weak-field, slow-motion behavior} 

The effects of the scalar field $\phi$ on the Solar 
System dynamics can be determined by studying the weak-field and slow-motion limit of 
Eqs.~(\ref{eq:evol-gab}) and (\ref{eq:evol-phi}). To do this, we consider an expansion of the 
metric and the scalar field about a cosmological solution, which sets the asymptotic boundary 
values, using a quasi-Minkowskian coordinate system, in which $g_{\mu\nu}\approx 
\eta_{\mu\nu}+h_{\mu\nu}$, with $|h_{\mu\nu}|\ll 1$. Denoting the asymptotic value of $\phi$ 
as $\phi_0$ and the local perturbation as $\varphi(x)$, to linear order Eq. 
(\ref{eq:evol-phi}) becomes
\begin{equation}\label{eq:linear-phi}
(\vec{\nabla}^2-m_\varphi^2)\varphi=\frac{\phi_0\kappa^2}{3}\rho \ ,
\end{equation}
where $m_\varphi^2 \equiv \left. (2V-V_{\phi}-\phi(1+\phi)V_{\phi\phi})/3\right|
_{\phi=\phi_0}$, and we have neglected the time derivatives of $\varphi$ (slow-motion regime). 
Imposing standard gauge conditions, the perturbations $h_{\mu\nu}=g_{\mu\nu}-\eta_{\mu\nu}$ 
satisfy the following equation
\begin{equation}\label{eq:linear-gab}
-\frac{1}{2}\vec{\nabla}^2h_{\mu\nu}=\frac{1}{1+\phi_0}\left(T_{\mu\nu}-\frac{1}{2}T
\eta_{\mu\nu}\right)+\frac{V_0+\vec{\nabla}^2\varphi}{2(1+\phi_0)}\eta_{\mu\nu} \ ,
\end{equation}
where to this order $T_{00}=\rho$, $T_{ij}=0$, $T=-\rho$. Far from the sources and assuming
spherical symmetry, the solution to Eqs.~(\ref{eq:linear-phi}) and (\ref{eq:linear-gab}) lead 
to ($M=\int d^3x \rho(x)$)
\begin{eqnarray}
\varphi(r)&=&\frac{2G}{3}\frac{\phi_0 M}{r}e^{-m_\varphi r} \label{cor1}\\
h_{00}^{(2)}(r)&=& \frac{2G_{\rm eff} M}{r} +\frac{V_0}{1+\phi_0}\frac{r^2}{6} \label{cor2} \\
h_{ij}^{(2)}(r)&=& \left(\frac{2\gamma G_{\rm eff} M}{r} -\frac{V_0}{1+\phi_0}\frac{r^2}
{6}\right)\delta_{ij}
\label{cor3}\ ,
\end{eqnarray}
where we have defined the effective Newton constant $G_{\rm eff}$ and the post-Newtonian 
parameter $\gamma$ as
\begin{eqnarray}
G_{\rm eff}&\equiv & \frac{G}{1+\phi_0}\left[1-\left(\phi_0/3\right)e^{-m_\varphi r}\right]\,, 
\label{g_eff}\\
\gamma &\equiv & \frac{1+\left(\phi_0/3\right)e^{-m_\varphi r}}{1-\left(\phi_0/3\right)e^{-
m_\varphi r}} \,.
\end{eqnarray}

As is clear from the above expressions, the coupling of the scalar field to the local system 
depends on the amplitude of the background value $\phi_0$. If $\phi_0$ is small, then $G_{\rm 
eff}\approx G$ and $\gamma\approx 1$ regardless of the value of the effective mass 
$m_\varphi^2$. When this mass squared becomes negative, the exponential terms in the above expressions become cosinus.
This contrasts with the result obtained in the metric version of $f(R)$ 
theories \cite{Olmo:2005jd}. In that case one finds 
$\varphi=\left(2G/3\right)\left(M/r\right)e^{-m_f r}$, $G_{\rm eff}\equiv  G
\left(1+e^{-m_f r}/3\right)/\phi_0 $, and $\gamma \equiv  \left(1-e^{-m_f r}/
3\right)/\left(1+e^{-m_f r}/3\right) $, which requires a large mass $m_f^2\equiv (\phi
V_{\phi\phi}-V_\phi)/3$ to make the Yukawa-type corrections negligible in local experiments. This may
be achieved in specific models implementing the chameleon mechanism \cite{Capozziello:2007eu}. \\
We note that the massless limit of the results derived  in this section are in complete agreement with the analysis of \cite{Damour-EFarese} for general massless scalar-tensor theories. \\

\section{Late-time cosmic speedup}

As a specific example of modified cosmological dynamics, we consider the spatially flat
Friedman-Robertson-Walker (FRW) metric
\begin{equation} \label{metric}
ds^2=-dt^2+a^2(t) d{\bf x}^2 \,,
\end{equation}
where $a(t)$ is the scale factor. The Ricci scalar is given by $R=6(2H^2+\dot{H})$, where 
$H=\dot{a}(t)/a(t)$ is the Hubble parameter, and $\dot{a}\equiv da/dt$. With this metric, Eq. 
(\ref{eq:evol-gab}) yields the following evolution equations
\begin{eqnarray}
3H^2&=& \frac{1}{1+\phi }\left[\kappa^2\rho +\frac{V}{2}-3\dot{\phi}\left(H+\frac{\dot{\phi}}
{4\phi}\right)\right] \ ,\label{field1} \\
2\dot{H}&=&\frac{1}{1+\phi }\left[ -\kappa^2(\rho+P)+H\dot{\phi}+\frac{3}
{2}\frac{\dot{\phi}^2}{\phi}-\ddot{\phi}\right] \ . \label{field2}
\end{eqnarray}

The scalar field equation (\ref{eq:evol-phi}) becomes
\begin{equation}
\ddot{\phi}+3H\dot{\phi}-\frac{\dot{\phi}^2}{2\phi}+\frac{\phi}{3}
[2V-(1+\phi)V_\phi]=-\frac{\phi\kappa^2}{3}(\rho-3P) \ .  \label{3}
\end{equation}
The qualitative behavior of the scalar field can be read directly from Eq. (\ref{3}) by 
writing the latter as follows
\begin{equation}
\ddot{\phi}+3H\dot{\phi}-\frac{\dot{\phi}^2}{2\phi}+M^2_\phi(T)\phi=0 \ ,  \label{3a}
\end{equation}
where $T=-(\rho-3P)$ and we have defined $M^2_\phi(T)$ as
\begin{equation}\label{eq:mass}
M^2_\phi(T)\equiv m_\phi^2-\frac{1}{3}\kappa^2T=\frac{1}{3}[2V-(1+\phi)V_\phi-\kappa^2T] \ ,
\end{equation}
which despite the notation needs not be a positive function. Aside from the 
$\dot{\phi}^2/\phi$ term, which is only important when $\phi$ is very rapidly changing, 
Eq.~(\ref{3a}) represents a massive scalar field on an FRW background.
During the matter dominated era ($T= -\rho$), the cosmic fluid contributes to the oscillation 
frequency of the scalar, while the friction term $3H\dot{\phi}$ forces a progressive damping 
of its amplitude. At late times, when $T \approx 0$, the sign of $m_\phi^2$ determines whether the field oscillates or grows exponentially 
fast. This aspect is model dependent and will be considered next. \\

\subsection{Two models}

We now propose two models that are consistent at Solar System and cosmological scales, and which 
are constructed on grounds of mathematical simplicity. A quantitative analysis of the high-precision astrophysical and cosmological data will 
be carried out elsewhere to find and constrain more general families of models within the 
metric-Palatini framework. 

The first model arises by demanding that matter and curvature satisfy the same relation as in 
GR. Taking
\begin{equation} \label{pot1}
V(\phi)=V_0+V_1\phi^2\,,
\end{equation}
Eq.~(\ref{eq:phi(X)}) automatically implies  
$R=-\kappa^2T+2V_0$. As $T\to 0$ with the cosmic expansion, this model naturally evolves into 
a de Sitter phase,  which requires $V_0\sim \Lambda$ for consistency with observations. If 
$V_1$ is positive, the de Sitter regime represents the minimum of the potential.  The 
effective mass for local experiments, $m_\varphi^2=2(V_0-2 V_1 \phi)/3$, is then positive and 
small as long as $\phi<V_0/V_1$. For sufficiently large $V_1$ one can make the field amplitude 
small enough to be in agreement with Solar System tests. It is interesting that the exact de Sitter solution
is compatible with dynamics of the scalar field in this model. 

A second model can be found by rewriting Eq.~(\ref{field1}) as
\begin{equation}\label{field1a}
\left(H+\frac{\dot{\phi}}{2(1+\phi)}\right)^2=\frac{\kappa^2\rho+V/2}
{3(1+\phi)}-\frac{\dot{\phi}^2}{4\phi(1+\phi)^2}
\end{equation}
and looking for late-time solutions, with 
\begin{equation}
H+\frac{\dot{\phi}}{2(1+\phi)}=\tilde{H}_0= {\rm constant},
\end{equation}
which leads to 
\begin{equation}
  a \sqrt{1+\phi}=a_0 e^{\tilde{H}_0t}.
\end{equation}  
 When $\rho\to 0$, the evolution equations and $\dot{\tilde{H}}_0=0$ lead to
\begin{equation}
\left(\tilde{H}_0^2-\frac{V}{6}\right)^2=9\tilde{H}_0^2\phi\left[\frac{V}
{6(1+\phi)}-\tilde{H}_0^2\right] \ ,
\end{equation}
from which one obtains
\begin{equation} \label{pot2}
V(\phi)=\frac{3\tilde{H}_0^2}{(1+\phi)}\left[2+11\phi\pm 3\phi \sqrt{5-4\phi}\right] \ .
\end{equation}
Remarkably, there are no free parameters in this model except for its amplitude 
$\tilde{H}_0^2$, which should be of the same order as the currently estimated cosmological 
constant, and the sign in front of the squared root. For the minus sign, we find that 
$m_\varphi^2$ is positive and of order $\sim\tilde{H}_0^2$ for  small $\phi$, which 
provides another model with a long-range scalar consistent with local tests and  
late-time cosmic speedup. 

\subsection{ Cosmological perturbations}

It should be noted that due to the appearance of an effective sound speed for matter 
perturbations, most pure Palatini-$f(R)$ models can be ruled out as dark 
energy candidates \cite{palatini_lss}. 
In the metric-Palatini approach presented here, however, such Laplacian instabilities are absent.  
Let us describe the relative perturbation of the matter distribution in a dust-dominated 
universe by $\delta \equiv \delta\rho/\bar{\rho}$, where $\bar{\rho}$ is the smooth background 
and $\delta\rho$ the inhomogeneous part. Allowing also the metric and the scalar field to 
fluctuate, one can consistently consider the evolution of perturbations. At subhorizon scales, 
that is described by
\begin{equation} \label{delta_evol}
\ddot{\delta} + 2H\dot{\delta} = 4\pi G_{eff}\bar{\rho}\delta\,, \quad  G_{\rm eff} \equiv 
\frac{1-\phi /3}{1+\phi}G\,.
\end{equation}
In accordance with (\ref{g_eff}), the effective Newton's constant becomes now time-dependent 
at cosmological scales. This modifies the growth rate of perturbations, and opens the 
possibility to distinguish between models with similarly accelerating background, e.g. \cite{Mota:2011iw}. In addition,
effective anisotropic stresses will appear, which could be detected in future weak lensing 
surveys such as the Euclid mission.

One may also verify that in vacuum the fluctuations $\delta\phi$ in the effective field $\phi$ at small scales
propagate with the speed of light,
Thus, despite the nonstandard coupling of the scalar-field, its perturbations behave physically. \\

\section{ Conclusions}

In this work we have considered a class of modified gravity actions where the corrections to 
the Einstein-Hilbert term are given by a function of the Ricci scalar constructed from an {\it 
a priori} metric-independent connection. We have shown that the theory admits a scalar-tensor 
representation which possesses a new mechanism to pass the Solar System constraints even if 
the scalar field is very light. This can be seen from the first order Post-Newtonian 
corrections to the metric given in Eqs.~(\ref{cor1})-(\ref{cor3}). If the current cosmic 
amplitude $\phi_0$ is sufficiently small, $|\phi_0|\ll 1$, a long-range scalar field able to 
affect the cosmic and galactic dynamics can also be compatible with the Solar System dynamics. 

Motivated by mathematical simplicity, we have presented two cosmological models with 
asymptotically de Sitter behavior. In both cases the early time evolution also seems 
consistent with the well-known radiation and matter dominated phases of the cosmic evolution. 
The potential (\ref{pot2}) had only one adjustable parameter, which is set by the 
observed scale of acceleration. The simpler potential (\ref{pot1}) included 
also the parameter fixing the magnitude of the quadratic correction. A quantitative analysis of 
structure formation and CMB anisotropies can thus be used to test the viability of this and other models. 

Before concluding, we point out that a modified gravitational potential of the form $\Phi =-
GM\left[ 1+\alpha _0\exp \left( -r/r_{0}\right) \right] /\left(1+\alpha _0\right) r$, with 
$\alpha _0=-0.9$ and $r_{0}\approx 30$ kpc \cite{Sa84} provides a very good description of the 
flat rotational curves of a significant sample of galaxies. The form of the weak-field 
potentials of the metric-Palatini hybrid model considered in this work share an interesting 
formal resemblance with this proposal. Thus, in addition to passing the Solar System 
constraints, the theory considered in this work may open new possibilities to approach, 
in the same theoretical framework, of both dark matter and dark energy problems. 
Also such aspects as future singularities \cite{Capozziello:2009hc} would be interesting
study in the hybrid theory.
Further work along these lines is presently underway.

{\it Acknowledgments}.
GJO is supported by the Spanish grant FIS2008-06078-C03-02 and the 
Consolider Programme CPAN (CSD2007-00042). The work of TH is supported by an RGC grant of the 
government of the Hong Kong SAR. FSNL acknowledges financial support of the Funda\c{c}\~{a}o 
para a Ci\^{e}ncia e Tecnologia through the grants PTDC/FIS/102742/2008 and 
CERN/FP/116398/2010.


\begin{thebibliography}{99}

\bibitem{fRgravity}
A.~De Felice and S.~Tsujikawa.
  Living Rev.\ Rel.\  {\bf 13}, 3 (2010);
  T.~P.~Sotiriou and V.~Faraoni, Rev. Mod. Phys. {\bf 82}, 451 (2010);
  S.~Nojiri and S.~D.~Odintsov,
  Phys.\ Rept.\  {\bf 505}, 59 (2011);
  %
  F.~S.~N.~Lobo,
  arXiv:0807.1640 [gr-qc].
  S.~Capozziello and M.~De Laurentis,
  Phys.\ Rept.\  {\bf 509}, 167 (2011)

\bibitem{Amendola:2010bk}
  L.~Amendola, K.~Enqvist, T.~Koivisto,
  Phys.\ Rev.\  {\bf D83}, 044016 (2011);
%
  T.~S.~Koivisto,
  Phys.\ Rev.\  D {\bf 83}, 101501 (2011);
%
  T.~S.~Koivisto,
  [arXiv:1109.4585 [gr-qc]].


\bibitem{Harko:2010mv}
  T.~Koivisto,
  Class.\ Quant.\ Grav.\  {\bf 23}, 4289-4296 (2006);
%
 T.~Harko and F.~S.~N.~Lobo,
  Eur.\ Phys.\ J.\  C {\bf 70}, 373 (2010);
%
 T.~Harko, T.~S.~Koivisto and F.~S.~N.~Lobo,
  Mod.\ Phys.\ Lett.\  A {\bf 26}, 1467 (2011);
%
  O.~Bertolami, C.~G.~Boehmer, T.~Harko and F.~S.~N.~Lobo,
  Phys.\ Rev.\  D {\bf 75}, 104016 (2007);
%
O. Bertolami and J. P\'aramos, Phys.\ Rev.\  D {\bf 77}, 084018
(2008);
%
  O.~Bertolami, F.~S.~N.~Lobo and J.~Paramos,
  Phys.\ Rev.\  D {\bf 78}, 064036 (2008);
%
T.~Harko, F.~S.~N.~Lobo, S.~Nojiri and S.~D.~Odintsov,
  Phys.\ Rev.\  D {\bf 84}, 024020 (2011).

\bibitem{expansion}
P. J. E. Peebles and B. Ratra, Rev. Mod. Phys. \textbf{75}, 559 (2003); T. Padmanabhan, Phys. 
Repts. \textbf{380}, 235 (2003).

\bibitem{Carroll:2003wy}
  S.~M.~Carroll, V.~Duvvuri, M.~Trodden and M.~S.~Turner,
  Phys.\ Rev.\  D {\bf 70}, 043528 (2004).

\bibitem{Olmo:2011uz}
  G.~J.~Olmo,
  Int.\ J.\ Mod.\ Phys.\  {\bf D20}, 413-462 (2011).

\bibitem{Flanagan:2003iw}
  E.~E.~Flanagan,
  Class.\ Quant.\ Grav.\  {\bf 21}, 417-426 (2003).

\bibitem{Olmo:2008nf}
  G.~J.~Olmo, P.~Singh,
  JCAP {\bf 0901}, 030 (2009).


\bibitem{Olmo:2005jd}
  G.~J.~Olmo, Phys.\ Rev.\ Lett. {\bf 95}, 261102(2005);  Phys.\ Rev.\  {\bf D72}, 083505 (2005).

\bibitem{Capozziello:2007eu} 
  S.~Capozziello and S.~Tsujikawa,
  Phys.\ Rev.\ D {\bf 77}, 107501 (2008)

\bibitem{Damour-EFarese}
T.Damour  and G. Esposito-Farese,  
Class. Quantum Grav., 9, 2093 (1992). 


\bibitem{palatini_lss}
  T.~Koivisto, H.~Kurki-Suonio,
  Class.\ Quant.\ Grav.\  {\bf 23}, 2355 (2006);
  T.~Koivisto,
  Phys.\ Rev.\  {\bf D73}, 083517 (2006).

\bibitem{Mota:2011iw} 
  D.~F.~Mota, V.~Salzano and S.~Capozziello,
  Phys.\ Rev.\ D {\bf 83}, 084038 (2011)

\bibitem{Sa84}  R. H. Sanders, Astron. Astrophys. {\bf 136},
L21 (1984); R. H. Sanders, Astron. Astrophys. {\bf 154}, 135
(1986).

\bibitem{Capozziello:2009hc} 
  S.~Capozziello, M.~De Laurentis, S.~Nojiri and S.~D.~Odintsov,
  Phys.\ Rev.\ D {\bf 79}, 124007 (2009)

\end{thebibliography}
\end{document}